\documentclass{article}
\usepackage{spconf,amsmath,graphicx}

\usepackage{subfigure}
\usepackage{booktabs}

\usepackage{threeparttable}

\PassOptionsToPackage{dvipsnames}{xcolor}
\usepackage{color}
\newcommand{\ie}{i.\,e.,\ }

\title{WakeupNet: A Mobile-Transformer based Framework \\ for End-to-End Streaming Voice Trigger}

\name{Zixing Zhang, Thorin Farnsworth, Senling Lin, Salah Karout} 
\address{Huawei Technologies Research \& Development (UK) Ltd \\ zixingzhang@huawei.com}

\ninept

\begin{document}
%
\maketitle
\begin{abstract}
End-to-end models have gradually become the main technical stream for voice trigger, aiming to achieve an utmost prediction accuracy but with a small footprint. In present paper, we propose an end-to-end voice trigger framework, namely \textit{WakeupNet}, which is basically structured on a Transformer encoder. The purpose of this framework is to explore the context-capturing capability of Transformer, as sequential information is vital for wakeup-word detection. However, the conventional Transformer encoder is too large to fit our task. To address this issue, we introduce different model compression approaches to shrink the vanilla one into a tiny one, called \textit{mobile-Transformer}. To evaluate the performance of mobile-Transformer, we conduct extensive experiments on a large public-available dataset HiMia. The obtained results indicate that introduced mobile-Transformer significantly outperforms other frequently used models for voice trigger in both clean and noisy scenarios.
\end{abstract}

\section{Introduction}
Nowadays, voice assistants have become increasingly popular in our daily life~\cite{Zhang17-Advanced}. 
In the systems, \textit{voice trigger} (\textit{aka} keyword spotting or wakeup-word detection) is considered as one of the frontier components, taking the responsibility to trigger the voice assistants so as to initialise the control or interaction process. Therefore, the prediction accuracy of voice trigger has a strong impact on the user experience of voice assistants. 
Besides, it is of significant importance to keep voice trigger system hardware-efficient as well due to its always-on characteristics. Thus, utmost reducing its storage and computational cost to fit the memory and energy constraint  is of necessary. 

Over the past few years, \textit{Transformer} encoder as well as its variants like Bert~\cite{Devlin19-BERT, Liu19-RoBerta,Sanh19-DistilBERT,Lan20-Albert},  have been widely used in natural language processing (NLP)~\cite{Liu19-RoBerta,Sanh19-DistilBERT,Lan20-Albert}. 
The major advantage of Transformer is its efficiency in extracting context-dependent representations~\cite{Vaswani17-Attention}. It can explicitly explore the context dependence  over a long sequence by a self-attention mechanism.  
Compared with Recurrent Neural Networks (RNNs), such as long short-term memory (LSTM) or Gated Recurrent Unit (GRU) RNNs, Transformer avoids the recurrent process which is considered to be unfriendly for parallel computation when using GPU. Thus, it largely facilitates the model training and inference processes.  

Encouraged by its great success in NLP, Transformer encoder has recently attracted increasing attention in other domains, such as computer vision~\cite{Girdhar19-Video} and speech processing~\cite{Chang20-End,Wang20-Transformer}. For example, in~\cite{Karita19-Comparative}, the authors conducted comprehensive studies on the comparison among Transformer, LSTM, and Kaldi-based hybrid models for automatic speech recognition (ASR), and found that the Transformer achieve the best performance on most datasets. 

However, the vanilla Transformer encoder was designed without considering deploying them in an edge device. This issue largely impedes its applications because these devices normally have strong storage and energy consumption limitations. Recently, much effort has been made toward compressing the model size. For example, DistilBert~\cite{Sanh19-DistilBERT} was introduced by distilling knowledge from a large model to a light one. 
In the context of voice trigger, nevertheless, these models are still far  larger than the ones we need. To this end, in this paper, we propose a compressed Transformer encoder, namely \textit{mobile-Transformer}, to enable the conventional model to fit the task of voice trigger. Besides, we take an end-to-end framework that is able to detect wakeup words streamingly. 


%
%
%
%
%
%
%
%

\section{Related Work}
\label{sec:relatedWork}
In the literature, many approaches were introduced for voice trigger, which can be grouped into \textit{filler-based}~\cite{Sun17-Compressed} and \textit{end-to-end} ones~\cite{Alvarez19-End,Rybakov20-Streaming}. The former approaches regard all background noise and non-wakeup speech as fillers, and model both the wakeup words and the fillers; whereas the later approaches model the offset of wakeup words and the others. 

Typical \textit{filler-based} approaches seek the help from ASR systems, where 
hidden Markov models (HMMs) are used to represent both the wakeup word (\textit{aka} keyword) and the background audio~\cite{Sun17-Compressed}.
However, its performance highly depends on its prediction accuracy of phoneme predictions. Besides, the complexity of ASR systems increases their deployment difficulty due to the high memory and power requirements. 
To overcome these issues, neural network-only based approaches were then proposed. They utilised advanced deep learning models to predict the wakeup words framewisely and straightforwardly by stacking multiple acoustic frames as inputs. Then, a sliding window is applied to average the posteriors. Once the smoothed value surpasses a pre-defined threshold, a wakeup word is supposed to be detected. Typical work can be found in~\cite{Sainath15-Convolutional,Arik17-Convolutional,Shan18-Attention},  where convolutional neural network (CNN), LSTM-RNNs, convolutional RNNs (CRNN) were applied respectively.

Nowadays, \textit{end-to-end} approaches have gradually become the mainstream technology for voice trigger, which straightforwardly estimate the wakeup point of keywords~\cite{Alvarez19-End}. Compared with the filler-based approaches, the end-to-end structure becomes simpler. Besides, it was investigated to be more effective as it directly optimises the detection score~\cite{Alvarez19-End,Rybakov20-Streaming}. Typical work can be found in~\cite{Alvarez19-End,Rybakov20-Streaming} where only the offset of wakeup words is annotated as positive. In this paper, we focus on the end-to-end detection system and a novel tiny model is designed and investigated. 

%
%
%
%
%
%
%

\section{Transformer-based End-to-End Architecture}
\label{sec:method}

In this section, we briefly introduce the WakeupNet framework for end-to-end  streaming voice trigger first. We then describe the proposed mobile-Transformer and employed focal loss. 

\subsection{Overview of WakeupNet Framework} 
\label{sec:framework}
\begin{figure}[!t]
	\centering
	\includegraphics[width=3.3in,trim={4.cm 0cm 6cm 0cm}, clip]{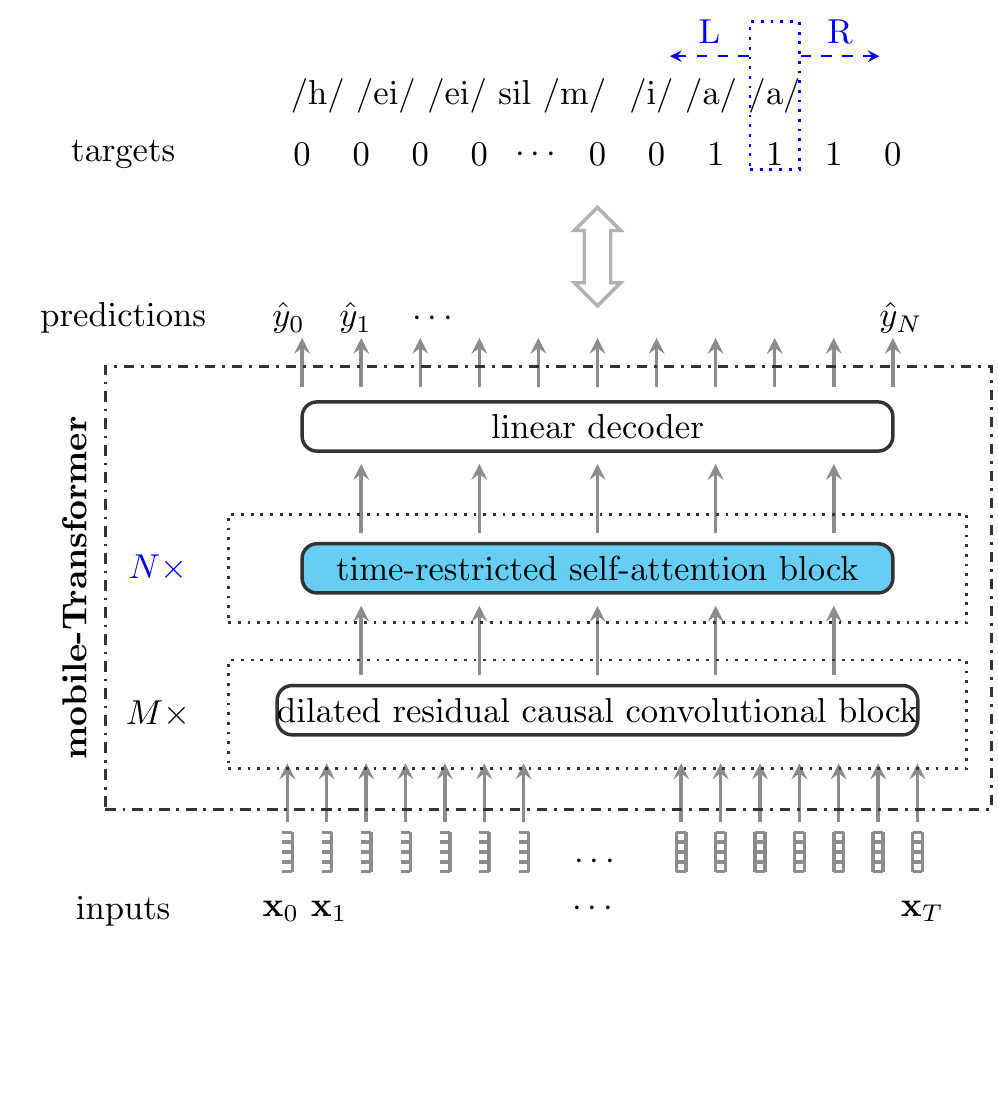}
	\vspace{-1.2cm}
	\caption{WakeupNet framework of the end-to-end streaming voice trigger based on a mobile-Transformer encoder.}
	\label{fig:framework}
\end{figure}

The WakeupNet framework for voice trigger is illustrated in Fig.~\ref{fig:framework} by using mobile-Transformer encoder as a backbone. Given sequential acoustic features $\{\mathbf{x}_t, t=0, \ldots, T\}$ and corresponding labels $\{y_i\in[0,1], i=0, \ldots, I\}$ where $T$ and $I$ are corresponding frame and label numbers respectively, the system aims to find a nonlinear mapping function $f$ between them. 

Regarding to the annotations, we followed the same principle in~\cite{Alvarez19-End} where only the end of the wakeup words is annotated as positive ($y_t=1$) and rest of them as negative ($y_t=0$). The benefits are at least twofold: i) it directly optimises the detection task and avoids any intermediate components compared with the filler-based systems~\cite{Alvarez19-End}; ii) it ultimately avoids the advanced and delayed problems when triggering the voice~\cite{Alvarez19-End}. 
Nonetheless, this annotation way leads a high data imbalanced problem, where the negative labels are far more than the positive labels. To deal with this, we repeated $L$ times before, and $R$ times after the original positive labels, such that the number of positive labels increases $L + R$ times to relieve the data imbalanced problem. 

In the inference stage, acoustic features are extracted from the streaming audio signals, and then are segmented into sequential clips $\{S_j, j=0, \ldots, J\}$ with the same WakeupNet perception field of $W$ frames and the step size of $P$ frames. Thus, generally speaking $I = J = \lfloor T/P\rfloor$. 
After that, a smoothed window is applied to the obtained sequential predictions  $\{\hat{y}_i\in[0,1], i=0, \ldots, I\}$. Once the smoothed predictions $\tilde{y}_i$ are higher than a pre-defined threshold score $s$, the system is triggered.

\subsection{Mobile-Transformer}

\begin{figure}[!t]
	\centering
	\includegraphics[width=2.5in,trim={5.3cm 8.5cm 6.5cm 0cm}, clip]{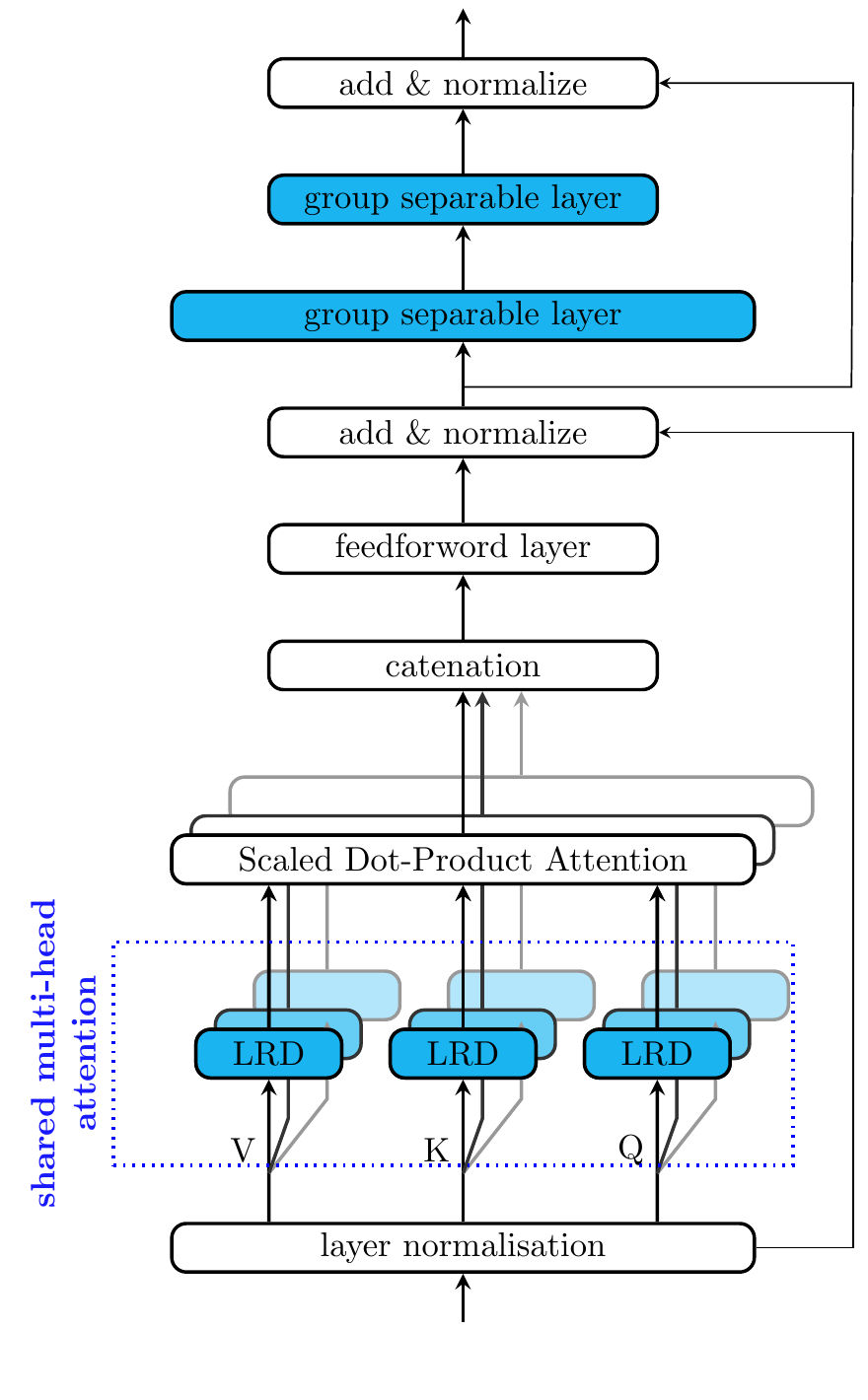}
	\vspace{-0.6cm}
	\caption{A diagram of the self-attention block by taking low-rank decomposed (LRD) attention layers, shared attention modules, and separable feed-forward layers.}
	\label{fig:mobileTrans}
\end{figure}

The introduced mobile-Transformer is composed with $M$ stacked VGG convolutional blocks~\cite{Simonyan15-Very}, $N$ stacked time-restricted self-attention blocks~\cite{Vaswani17-Attention}, and a linear decoder, as illustrated in Fig.~\ref{fig:mobileTrans}. 

For sequence encoding, the position information of its elements is of importance. However, the attention mechanism in Transformer is neither recurrent nor convoluted~\cite{Vaswani17-Attention}. In NLP, one simple way is adding a position encoding~\cite{Vaswani17-Attention}. Different from this explicit way, in speech processing domain an implicit way have shown to be efficient by using convolutional operations~\cite{Chorowski15-Attention, Yeh19-Transformer}, which is supposed to automatically capture the contextual information when a deep structure is taken~\cite{Yeh19-Transformer}. Besides, for streaming inference, causal convolution is considered such that only historical signals need to be collected for inference. More details of the VGG blocks can be found in~\cite{Vaswani17-Attention}. 

As to the self-attention blocks, it contains a stack of multiple identical blocks. Each block contains an attention layer and two feedforward layers. The attention layer first applies LayerNorm, then projects the input to queries ($Q$), keys ($K$), and values ($V$). The attention output is calculated by 
 \begin{equation}\label{eq:att}
 \mathtt{Attention}(Q, K, V) = \mathtt{softmax}(\frac{QK^T}{\sqrt{d_k}})V, 
 \end{equation}
where $d_k$, the dimension of keys, is the scaled factor. In doing this, each obtained representation implicitly contains the semantic information over the whole sequence.  
To jointly attend the information from different representation subspaces at different positions, we apply the multihead strategy~\cite{Vaswani17-Attention} as well by splitting queries, keys, and values into several parts. After that, two feedforward layers with ReLU activation functions follow to increase the non-linear learning capability of the blocks. 
For the self-attention layers and feedforward layers, residual adding and layer normalisating are applied to deal with the gradient vanishing problem when increasing the depth of networks and the internal covariate shift problem, respectively. 

As mentioned in Section I, low latency and computation is vital for voice trigger due to its always-on nature. Therefore, for the Transformer encoder blocks, we took the suggestions in~\cite{Yeh19-Transformer}, where truncated self-attention is utilised because of i) its capability of streaming inference; i) its efficiency of computational complexity. Compared with the original self-attention that depends on the entire input sequence $\{\mathbf{x}_t, t=0, \ldots, T\}$, the truncated one only accesses the sequence $\{\mathbf{x}_t, t=t-h, \ldots, t, \ldots, t+b\}$ at time $t$, with $h$ look-ahead and $b$ look-back.



The vanilla Transformer encoder is shown in~\cite{Vaswani17-Attention}, where the weights mainly come from attention layers, feedforward layers, as well as its stacked structure. In the following, we compress the model in three ways. 

\subsubsection{Cross Layers Parameter Sharing}
Encouraged by the success of ALBERT -- a lite BERT structure~\cite{Lan20-Albert}, 
a cross-layer parameter sharing is employed for mobile-Transformer, where only the attention parameters are shared cross layers. We do neither take the all-shared strategy or the feedforward-shared strategy since they were empirically demonstrated to degrade the model performance greatly~\cite{Lan20-Albert}. 
The motivation of the cross layer parameter sharing is that the semantic relationship among the sequence is supposed to be similar although in different layers. 
By doing this, the number of attention weights can be significantly reduced to $1/N$ of its original size~\cite{Lan20-Albert}. 

\subsubsection{Low-Rank Decomposition}
According to Eq.~(\ref{eq:att}), each scaled dot-product attention layer contains three $d\times d$ attention weights, which contributes greatly to the entire model size. To compress the attention matrices, we consider low-rank decomposition (LRD). 
LRD maps high-dimensional vectors into a low-rank sub-space, and then reconstruct them into another high-dimensional vector with minimum information loss. Although LRD has been widely used for matrix compression to save storage and reduce computational complexity, its effectiveness in the context of Transformer has yet to be investigated, to the best of our knowledge. 

In present work, we insert a bottleneck layer between the input and output layers to simulate LRD like the work in~\cite{Xue14-Singular}, such that the number of attention weights $d\times d$ become $2 \times d\times r$. Thus, value of $r$ determines the compression rate $d/2r$. That is, when $r=d/4$, the matrix size is then reduced by half. 


\subsubsection{Group Separable Convolution} 
As the proposed mobile-Transformer does not share the feedforward layers cross transformer blocks, the feedforwards layers become another major weight contributor. To shrink this component, we make use of group convolution and separable convolution. 
Group convolution and separable convolution, which were firstly proposed in AlexNet~\cite{Krizhevsky12-ImageNet} and MobileNet~\cite{Howard17-MobileNets} respectively, are considered as two alternative convolutional ways. 
In group convolution, the filters are separated into different groups. Each group is responsible for conventional convolutions with certain depth~\cite{Zhang18-ShuffleNet}. The group convolution is efficient in training as i) it shrinks the original convolution into $1/g$, where $g$ is the number of groups; ii) it allows the model training over multiple GPUs in a parallel fashion. Besides, each filter group is able to learn a unique representation of the data~\cite{Zhang18-ShuffleNet}. 

The separable convolution normally contains two steps: i) a depthwise convolution, where a single filter per each input channel is applied, and ii) a pointwise convolution, where a simple 1×1 convolution is then used to create a linear combination of the output of the depthwise layer~\cite{Howard17-MobileNets}. The ratio of the learnable weights between the depthwise separable convolution and the 2D convolution can be expressed as~\cite{Howard17-MobileNets}: 
\begin{equation}
\frac{1}{C_o} + \frac{1}{D_{K}^{2}}, 
\end{equation}
where $C_o$ is the number of output channels, and $D_k$ denotes the spatial dimension of the depthwise kernel assumed to be square.  
Note that, in present work, the 2D operation is down to 1D as the inputs of feedforward layers are vectors only.

\subsection{Focal Loss} 

To train mobile-Transformer, we choose focal loss as the objective function~\cite{Lin20-Focal} due to its effectiveness when training with unbalanced data distribution over class. 
Mathematically, the focal loss is formulated by 
\begin{equation}\label{eq:fl}
\text{FL}(p_t)=-\alpha_t(1-p_t)^\gamma \text{log}(p_t), 
\end{equation}
where 
\begin{equation}
p_t = y * \hat{y} + (1 - y) * (1 - \hat{y}). 
\end{equation}
The hyperparameter $\alpha\in[0,1]$ controls the importance of positive/negative samples in a linear way, whilst does not make any difference between easy/hard samples. By contrast, $\gamma\geq0$ controls the importance of easy/hard samples in an exponential way, whilst does not make any difference between positive/negative samples. Higher $\gamma$ forces the model to learn more from difficult (hard) samples. When $\alpha$ is set to be 0.5 and $\gamma$ to be 0.0, the focal loss is equal to conventional cross entropy. More details about focal loss can be found in~\cite{Lin20-Focal}.

\section{Experiments and Results} 
\label{sec:experiments}
In this section, we describe the selected dataset HiMia, experimental setups and results, and provide a discussion. 

\subsection{Datasets} 
To evaluate mobile-Transformer for voice trigger, we selected the Mandarin Chinese dataset HiMia~\footnote{http://www.openslr.org/85} (wakeup word: `ni hao, mi ya')~\cite{Qin20-Himia} due to the large data size it has and the variety of acoustic environments it covers.  Basically, the dataset includes two subsets: AISHELL-wakeup and AISHELL-2012B-eval, which cover 254 and 82 speakers respectively, and contain 995M and 330M samples respectively. The average sample duration is around 1.5 second. For our task, AISHELL-wakeup was used as the training set and AISHELL-2012B-eval was split into development and test sets. The detailed data distribution is shown in Table~\ref{tab:himia}. More importantly, the recordings were collected in both close and far-field scenarios, in both acoustic clean and noisy scenarios, and in different speech speed (i.e., slow, normal, and fast). All these variations significantly reduce the data mismatch for model training and in real-life application. 

\begin{table}[!t]
	\centering
	\caption{Data distribution of HiMia dataset over three partitions.}
	\begin{threeparttable}
		\begin{tabular}{lrrr}
			\toprule
			HiMia & \# speakers & \# samples & hours \\ 
			\midrule 
			train & 254 & 995,680 & 417 \\ 
			dev & 42 & 164,640 & 61 \\ 
			test & 44 & 165,120 & 46 \\
			total & 340 & 1,325,440 & 524 \\
			\bottomrule			
		\end{tabular}
	\end{threeparttable}
	\label{tab:himia}
\end{table}

For negative samples, we chose another Mandarin Chinese dataset MAGICDATA~\footnote{http://www.openslr.org/68}, which contains 755 hours of scripted read speech data from 1\,080 native speakers. Its training, development and test sets include 573\,480, 11\,793, and 24\,279 samples, respectively. 
Besides, to conduct audio augmentation, we took the frequently used MUSAN noisy dataset~\cite{Snyder15-Musan}, which includes music, speech, and noise recordings. 

\subsection{Experimental Setups} 
We used two VGG blocks ($M=2$) and three ($N=3$) time-restricted self-attention blocks for mobile-Transformer. Each VGG block contains one convolutional layer with the kernel size (3, 3) and 32 channels, which is then followed by a (2, 2) max-pooling to reduce the feature dimension and increase the network perception field. 
As to the self-attention blocks, 
the dimension $d$ of attention and  $r$ of their corresponding low-rank layer were set to be 64 and 16, respectively. For the group separable convolution, the number of groups was set to be 4. All these hyperparameters were optimised by an empirical model evaluation on the development set, taking the model size and performance into account. In total, the wakeup model has 85\,K parameters. Additionally, to control the model latency, we set the look-ahead frames $h$ to be 0. 

To increase the positive annotation numbers, we repeated six times before and after the expected wakeup point ($L=R=6$). Due to two max-pooling layers, the length of annotation sequence is reduced to 1/4 of the input sequence ($P=4$). To train mobile-Transformer, Adam optimiser and a warmup-hold-exponential\_decay learning rate schedule (start, hold, and final learning rates: $10^{-6}$, $5*10^{-5}$, $10^{-6}$) was used to increase the learning efficiency and stability. For the focal loss, the $\alpha$ and $\gamma$ were optimised to be 0.1 and 1 in order to upweight the loss contribution from positive and hard samples, respectively. 

To evaluate the mobile-Transformer, we selected other frequently used voice trigger models for performance comparison. These models include  CNN~\cite{Sainath15-Convolutional}, RNN~\cite{Shan18-Attention}, ResNet~\cite{Tang18-Deep}, and CRNN~\cite{Arik17-Convolutional}. 
More specifically, we took the cnn-tstride4 architecture in~\cite{Sainath15-Convolutional} for CNN, the same RNN and CRNN architectures in~\cite{Shan18-Attention} and~\cite{Arik17-Convolutional} for RNN and CRNN, and the res8-narrow architecture in~\cite{Tang18-Deep} for ResNet in our experiments. Nevertheless, we slightly trimmed the original version into a smaller one, for the sake of a fair performance comparison. For example, the channel numbers of cnn-tstride4 were rescaled into 70 and 48; and the channel number of res8-narrow was shrunk into 32. 


%
%

%
%
%

\subsection{Experimental Results}
Table~\ref{tab:vs_models} compares the performance in terms of False Rejection Rate (FRR) at 10.0 and 0.5 False Alarm per Hour (FApH) between  mobile-Transformer and other frequently used voice trigger models (\ie CNN, RNN, ResNet, and CRNN) in both clean and mixed noisy scenarios. Particularly, the mixed noisy test data were simulated by adding MUSAN noise with SNRs in the range of [0dB, 20dB] and a step size of 5dB. With a similar model size, it can be seen from Table~\ref{tab:vs_models} that  mobile-Transformer achieves the best performance in terms of 0.5 FApH under the clean scenario (\ie 1.29\,\% FRR), and significantly (one-tailed $z$-test, $p<.05$) outperforms the second best model (CRNN). Such a performance improvement is even boosted in the noisy scenario (2.43\,\% FRR by mobile-Transformer vs 4.02\,\% FRR by ResNet), which implies that the mobile-Transformer shows much more noise-robust compared with other models. This improvement possibly attributes to the representation learning capability of mobile-Transformer. 

\begin{figure}[!t]
	\centering
	\includegraphics[width=1.55in,trim=0cm 0 1.0cm 0cm]{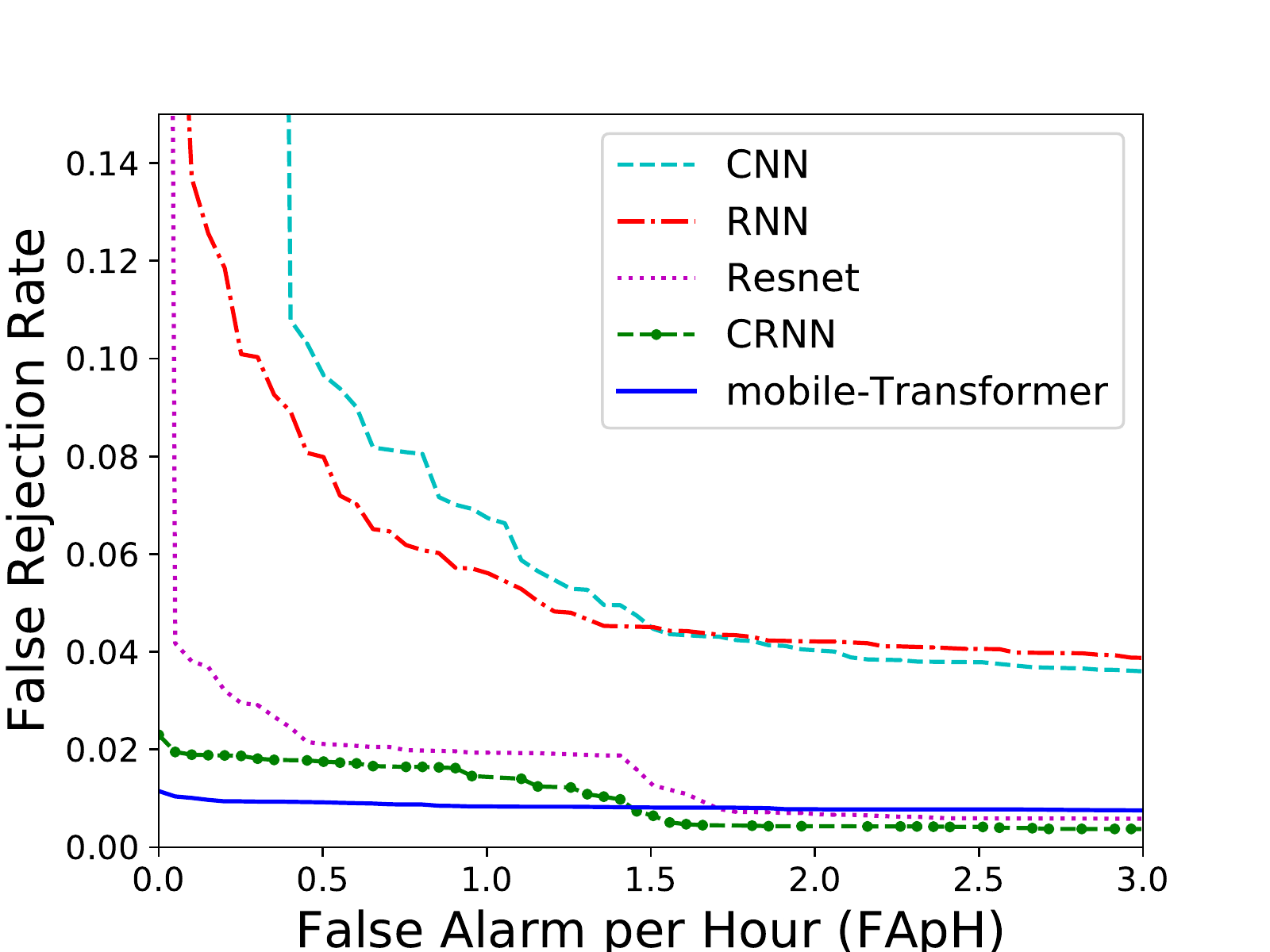} 
	\includegraphics[width=1.55in,trim=0cm 0 1.0cm 0cm]{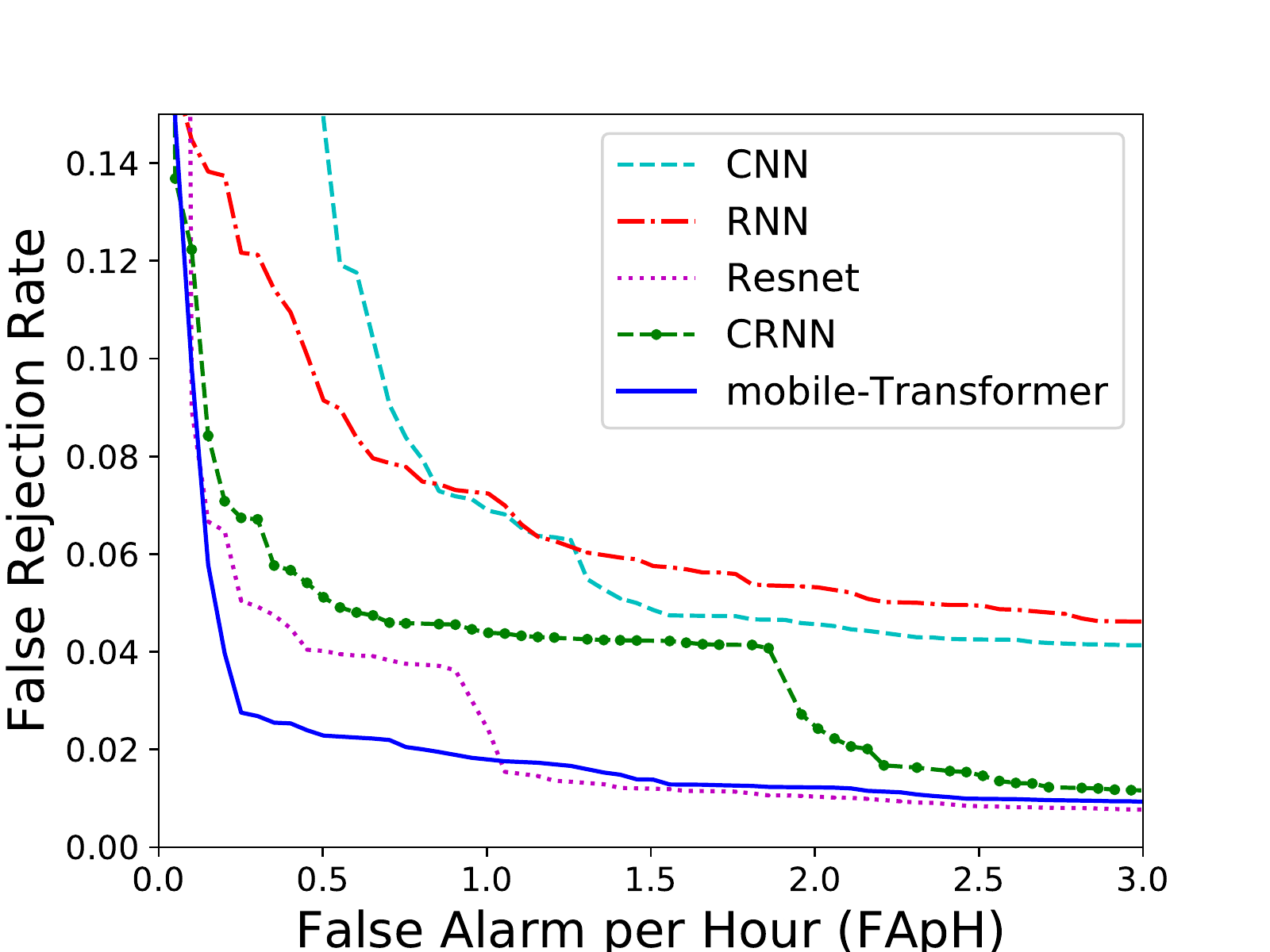} \\
	{(a) clean \quad \quad \quad \quad \quad \quad \quad \quad \quad \quad (b) noisy}
	\caption{Detection Error Trade (DET) curves obtained in the clean (a) and mixed noisy (b) scenarios on HiMia dataset.}
	\label{fig:det}
\end{figure}

To better show the trade-off between FRR and FApH, we plot the Detection Error Trade (DET) curves in Fig.~\ref{fig:det} for five compared models. Again, the proposed  mobile-Transformer obviously performs best among the five models at 0.5 FApH in either clean or noisy scenario, whereas the CRNN, Resnet, and  mobile-Transformer models perform competitively at high FApH values. 

\begin{figure}[!t]
	\centering
	\includegraphics[width=3.3in, trim={0cm 0cm 1.6cm 1cm}, clip]{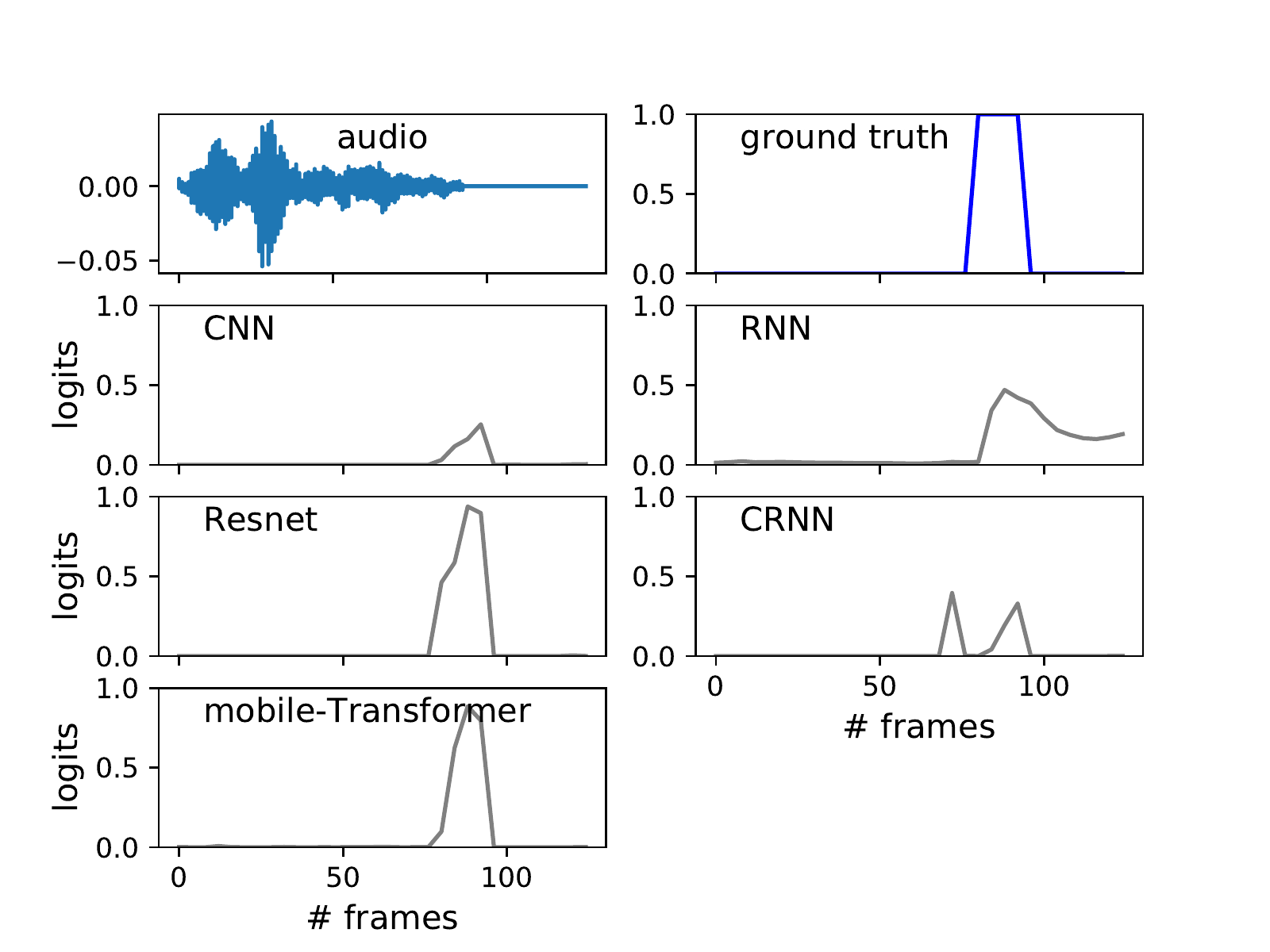} 
	\vspace{-.2cm}
	\caption{Prediction curves via different models.}
	\label{fig:predictions}
	\vspace{-.6cm}
\end{figure}

We further illustrated the frame-wise predictions on a randomly selected audio by using all compared models in Fig.~\ref{fig:predictions}. Generally speaking, one can see that the predictions from ResNet and  mobile-Transformer are closer to the annotations than other models. However, for CNN the prediction scores are not as high as the others although it correctly detects the wakeup point. This is possible due to its limited perception field as it is not deep enough. This conclusion is confirmed in ResNet that is much deeper than the employed CNN. Moreover, for RNN it is also able to predict the wakeup point accurately. But, the historical information over sequence greatly impacts its future predictions, as we observe that its predictions are not as sharp as the ones from non-recurrent models.

\begin{table}[!t]
	\centering
	\caption{Performance comparison between the proposed mobile-Transformer and other state-of-the-art models for voice trigger in clean and mixed noisy scenarios.} 
	\vspace{-.cm}
	\begin{threeparttable}
		\begin{tabular}{lrrrrrr}
			\toprule
			models & \# params & \multicolumn{2}{c}{0.5 FApH} \\
			\cmidrule{3-4} 
			FRR [\%] & & clean & noisy \\ 
			\midrule 
			CNN~\cite{Sainath15-Convolutional}	&	109K	&		9.69	&	15.55	\\
			GRU~\cite{Shan18-Attention}	&	97K	&	7.99	&	9.19	\\
			ResNet~\cite{Tang18-Deep}	&	88K	&	2.11	&	4.02	\\
			CRNN~\cite{Arik17-Convolutional}	&	114K	&	1.75	&	5.25	\\
			\midrule
			 mobile-Transformer	&	85K	&	\textbf{1.29}	&	\textbf{2.43}	\\
			\bottomrule
		\end{tabular}
	\end{threeparttable}
	\label{tab:vs_models}
\end{table}


\subsection{Discussions}


To investigate the impact of the perception field of mobile-Transformer, we compared the model performance when setting look-back frames $b$ to be 40, 60, and 80. The results are shown in Table~\ref{tab:vs_lookback}. On average, when $b=60$ it performs best in either clean or noisy scenario. This demonstrates that the small perception field is not able to cover the whole wakeup word, and the very large perception field takes much non-related noise which indeed increases the learning difficulty. Therefore, it is of importance to choose an appropriate perception field. 

We further evaluated the robustness of mobile-Transformer when re-scaling the model into larger or smaller ones. The results are displayed in Table~\ref{tab:vs_modelsize}. When comparing the mobile-Transformer models with vanilla or compressed Transformer encoder, we notice that larger the models are, the better performance they deliver. Nevertheless, when we compare the models with or without the proposed compression methods on the Transformer encoder, we found that the compressed model performance does not degrade. In contrast, they performs even better in most cases. This suggests that the combination of shared attention, LRD, and group separable convolution operations is able to efficiently reduce the model redundancy while keeping the model effectiveness. 

\begin{table}[!t]
	\centering
	\caption{Performance of mobile-Transformer when using different look-back frames per attention layer.} 
	\vspace{-.0cm}
	\begin{threeparttable}
		\begin{tabular}{lrrrrrr}
			\toprule
			\# frames & \multicolumn{2}{c}{10 FApH} & & \multicolumn{2}{c}{0.5 FApH} \\
			\cmidrule{2-3} \cmidrule{5-6} 
			FRR [\%] & clean & noisy && clean & noisy \\ 
			\midrule 
			40	&	0.82	&	0.97	&&	1.50	&	2.47	\\
			60	&	0.61	&	0.48	&&	1.29	&	2.43	\\
			80	&	0.80	&	0.76	&&	1.03	&	2.35	\\
			
			\bottomrule
		\end{tabular}
	\end{threeparttable}
	\label{tab:vs_lookback}
\end{table}

\begin{table}[!t]
	\centering
	\caption{Performance of the mobile- and vanilla-Transformer encoder in different model sizes when evlauated in clean and mixed noisy scenarios.} 
	\vspace{-.cm}
	\begin{threeparttable}
		\begin{tabular}{lrrrrrr}
			\toprule
			models & \# params & \multicolumn{2}{c}{10 FApH} & & \multicolumn{2}{c}{0.5 FApH} \\
			\cmidrule{3-4} \cmidrule{6-7} 
			FRR [\%] & & clean & noisy && clean & noisy \\ 
			\midrule 
			\multicolumn{6}{l}{\textit{vanilla Transformer encoder}}	\\								
			\quad	large	&	634K	&	0.75	&	0.78	&&	1.15	&	0.83	\\
			\quad	media	&	160K	&	0.36	&	1.15	&&	1.54	&	4.55	\\
			\quad	tiny	&	41K	&	0.82	&	1.71	&&	5.33	&	8.81	\\
			\multicolumn{6}{l}{\textit{mobile-Transformer encoder}}	\\								
			\quad	large	&	331K	&	0.68	&	0.91	&&	0.70	&	0.94	\\
			\quad	media	&	85K	&	0.61	&	0.48	&&	1.29	&	2.43	\\
			\quad	tiny	&	22K	&	0.65	&	1.20	&&	1.95	&	4.74	\\
			\bottomrule
		\end{tabular}
	\end{threeparttable}
	\label{tab:vs_modelsize}
\end{table}


%
\section{Conclusion} 
\label{sec:conclusion}

The capability of Transformer has been largely observed in many domains over the past few years, due to its efficient learning ability of context-dependent representations. However, it is unclear how it performs in voice trigger. To uncover this, we proposed WakeupNet -- a mobile-Transformer encoder based  end-to-end framework for streaming voice trigger. As it is essential to reduce model size, we compress the encoder by sharing attention weights cross layers, decomposing the attention weights into low ranks, and taking group separable  convolution operations as an alternative of feedforward connections. 

When compared with other state-of-the-art models on HiMia dataset, mobile-Transformer shows to be more effective in prediction accuracy at 0.5 False Alarm per Hour and to be more robust in noisy scenario. Besides, it is found that the introduced model compression approaches have little negative impact on its performance. 

Future work will focus on extracting representations on large-scale unlabelled data, to make it more efficient for low-resource wakeup words.

\bibliographystyle{IEEEbib}
\bibliography{mybib}

\end{document}